\documentclass[ALICE,manyauthors]{cernphprep}
\pdfoutput=1 
\usepackage{todonotes}
\usepackage[comma,square,numbers,sort&compress]{natbib}
\usepackage{hyperref}
\usepackage{lineno}
\usepackage{theorem}
\usepackage{color}
\usepackage{rotating}
\usepackage{epstopdf}
\usepackage{epsfig}
\usepackage{notoccite}
\usepackage{multirow}
\usepackage{tikz}
\usepackage{array}
\usepackage[font=small,labelfont=bf]{caption}
\usepackage{ltablex}
\usepackage[Symbol]{upgreek}
\usepackage{titlesec}
\usepackage{siunitx}
\usepackage{comment}

\hypersetup{
    colorlinks,
    citecolor=black,
    filecolor=black,
    linkcolor=black,
    urlcolor=black
}

\begin{document}

\begin{titlepage}
\PHyear{2019}
\PHnumber{045}      
\PHdate{13 March}  
%

\title{Measurement of strange baryon--antibaryon interactions with femtoscopic correlations}
\ShortTitle{Measurement of strange baryon--antibaryon interactions}   

\Collaboration{ALICE Collaboration\thanks{See Appendix~\ref{app:collab} for the list of collaboration members}}
\ShortAuthor{ALICE Collaboration} 

\begin{abstract}
Two-particle correlation functions were measured for $\rm p\overline{p}$, $\rm p\overline{\Lambda}$,  $\rm \overline{p}\Lambda$, and $\Lambda\overline{\Lambda}$ pairs in Pb--Pb collisions at $\sqrt{s_{\rm NN}}=2.76$~TeV and $\sqrt{s_{\rm NN}}=5.02$~TeV recorded by the ALICE detector. From a simultaneous fit to all obtained correlation functions, real and
imaginary components of the scattering lengths, as well as the effective ranges, were extracted for combined $\rm p\overline{\Lambda}$ and $\rm \overline{p}\Lambda$ pairs and, for the first time, for $\Lambda\overline{\Lambda}$ pairs. Effective averaged scattering parameters for heavier baryon--antibaryon pairs, not measured directly, are also provided. The results reveal similarly strong interaction between measured baryon--antibaryon pairs, suggesting that they all annihilate in the same manner at the same pair relative momentum $k^{*}$. Moreover, the reported significant non-zero imaginary part and negative real part of the scattering length provide motivation for future baryon--antibaryon bound state searches.
\end{abstract}
\end{titlepage}
\setcounter{page}{2}
\section{Introduction}
The interaction of baryons is a fundamental aspect of many sub-fields of nuclear physics. It is investigated extensively with numerous methods, among which are included the detailed analysis of the properties of atomic nuclei, the dedicated experiments where beams of one baryon type are scattered on other baryons bound in atomic nuclei~\cite{Stoks:1993tb,pap54,Eisele:1971mk,Hepp:1968zza,pap58,Bart:1999uh}, and the femtoscopy technique~\cite{Lednicky:1981su}. The latter involves the analysis of momentum correlations of two particles produced in nuclear or elementary collisions~\cite{Kopylov:1972qw, Kopylov:1973qq, Kopylov:1974th,Lednicky:2005af,Lisa:2005dd}. It is especially interesting to probe the interaction in the region of the low relative momentum of the pair, as it is the most relevant for a precise extraction of the strong interaction scattering parameters. In particular, the possible creation of bound states for a given baryon--baryon pair was investigated extensively~\cite{Hashimoto:2006aw,Hayano:1988pn,PhysRevLett.80.1605,Nakazawa:15,Jaffe:1976yi,Takahashi:2001nm}.

Nuclear collisions at relativistic energies are abundant sources of various particle species. In particular, the number of baryons and antibaryons created in each central Pb--Pb collision at the Large Hadron Collider (LHC)~\cite{Evans:2008zzb} is of
the order of one hundred each at mid-rapidity ($|y|<0.5$)  ~\cite{Abelev:2013xaa,Abelev:2012wca,Abelev:2013vea}, which makes it feasible to study details of their interactions. These particles include $\Lambda$, $\Sigma$, $\Xi$, and $\Omega$ and an approximately equal amount of their corresponding antiparticles.

The interactions of baryons are well known for pp pairs and pn pairs. Measurements were also performed for $\rm p\Lambda$ pairs~\cite{Alexander:1969cx,SechiZorn:1969hk,Kadyk:1971tc}. Recently, a comparative study of the baryon--baryon and antibaryon--antibaryon interaction using Au--Au collisions at $\sqrt{s_{\rm NN}}=200$~GeV has been performed by the STAR experiment at the Relativistic Heavy Ion Collider and found that the $\rm \overline{p}\overline{p}$ interaction does not differ from the pp system~\cite{Adamczyk:2015hza}.  Also, correlation measurements of baryon--baryon pairs in pp collisions at $\sqrt{s}=7$~TeV and p--Pb collisions at $\sqrt{s_{\rm NN}}=5.02$~TeV performed by the ALICE detector~\cite{Aamodt:2008zz} at the LHC provide more constraints on the interaction of $\rm p\Lambda$ and $\rm \Lambda\Lambda$~\cite{Acharya:2018gyz,Acharya:2019yvb} as well as $\rm p\Xi^{-}$~\cite{Acharya:2019sms} at low relative pair momentum.

Concerning proton--antiproton pairs, the strong interaction was studied in detail~\cite{Patrignani:2016xqp,Batty:1989gg,Pirner:1991mn,Grach:1997mc,Klempt:2002ap}. Of particular interest is \emph{protonium} (or \emph{antiprotonic hydrogen}) -- a proton--antiproton Coulomb bound state, where the strong interaction also plays a significant role. The protonium atoms are created by stopping antiprotons in hydrogen and the strong interaction is studied via shifts in the X-ray spectrum from the expected QED transitions from excited states. In particular, there is evidence of a contribution from the strong force to the 1S and 2P states. However, the nature of protonium in these states, whether it can be considered a nuclear bound state or a result of the Coulomb interaction, remains an open question. For more details we refer the reader to the review paper~\cite{Klempt:2002ap}.

For baryon--antibaryon pairs with non-zero strangeness there is much less experimental data available. However, low mass enhancements in the invariant mass distributions of $\rm p\overline{p}$, $\rm p\overline{\Lambda}$, and $\Lambda\overline{\Lambda}$ pairs have been observed in charmonium and $\rm B$ meson decays~\cite{Yuan:2005pf,Bai:2003sw,Ablikim:2004dj,Ablikim:2017pyl}. Those enhancements, except for the $\rm p\overline{p}$ pair, are slightly above the mass threshold of the baryon--antibaryon systems and have widths which are below 200~MeV/$c^2$. Theoretical interpretations of these results predict the existence of various baryon--antibaryon bound states and propose their classification~\cite{Yuan:2005pf}. Results presented in this letter might shed new light on this domain.

The baryon--antibaryon scattering parameters, when measured, could be implemented in the well-est\-ablish\-ed model of heavy-ion collisions, UrQMD~\cite{Bleicher:1999xi}, which has the important feature of including rescattering in the hadronic phase. In particular, recent comparisons of theoretical calculations with the ALICE data show that a proper description of this phase is critical for the correct reproduction of a large number of observables, like particle yields, transverse-momentum spectra, femtoscopy of identified particles, as well as elliptic flow~\cite{Steinheimer:2012rd,Steinheimer:2017vju,Werner:2012xh,Karpenko:2012yf}. The baryon--antibaryon annihilation is a critical component of the rescattering process. Yet, at the moment, for all but nucleon-antinucleon pairs, one has to rely on assumptions about the interaction cross section. Currently it is assumed that all baryon--antibaryon pairs annihilate in the same way as $\rm p\overline{p}$ pairs at the same total energy of the pair, $\sqrt{s}$, in the pair rest frame~\cite{Bleicher:1999xi}.

Femtoscopy allows one to access the baryon--antibaryon interaction at low pair relative momentum in a way which is complementary to dedicated scattering experiments. Only the strong interaction is present for $\rm\mathrm{p}\overline{\Lambda}$ ($\rm\overline{p}\Lambda$) and  $\rm \Lambda\overline{\Lambda}$ pairs, while for $\rm p\overline{p}$ pairs, where also the Coulomb interaction is present, it is the dominant contribution~\cite{Lednicky:1981su,Lednicky:2005tb}. Therefore, the parameters of this interaction, together with the source function, determine the shape of the correlation function. In addition, the so-called ``residual correlation'' effect (presence of an admixture of weak decay products in the sample of a given baryon--antibaryon pair) results in non-trivial interconnections between measured correlation functions. The femtoscopic technique has been employed already to measure $\rm p\overline{\Lambda}$ and $\rm \overline{p}\Lambda$ scattering parameters by the STAR experiment~\cite{Adams:2005ws}. However, the most important limitation of that study is the fact that no corrections for residual correlations were applied. 

In this letter the scattering parameters are extracted for $\rm p\overline{\Lambda}$, $\rm \overline{p}\Lambda$, and for the first time for $\Lambda\overline{\Lambda}$ pairs from femtoscopic correlations measured in Pb--Pb collisions at $\sqrt{s_{\rm NN}}=2.76$~TeV and $\sqrt{s_{\rm NN}}=5.02$~TeV registered by the the ALICE experiment at LHC. The residual correlations are accounted for in the formalism proposed in Ref.~\cite{Kisiel:2014mma} which does not attempt to ``correct'' for this effect (as proposed in an alternative procedure in Ref.~\cite{Shapoval:2014yha}), but instead uses it to extract information about the strong interaction potential parameters for the parent particles. Therefore, it allows for a single and simultaneous fit to all measured correlation functions. This provides maximum statistical accuracy for the obtained parameters, minimises the number of fit parameters and provides a non-trivial internal consistency verification. 

Recently, the $\rm p\overline{\Lambda}$ and $\rm \overline{p}\Lambda$ correlations measured by STAR~\cite{Adams:2005ws} have been reanalysed taking into account the residual correlations effect~\cite{Kisiel:2014mma}. That study suggests that all baryon--antibaryon pairs might annihilate in a similar way as a function of the relative momentum of the pair $k^{*}$, instead of the pair centre-of-mass energy $\sqrt{s}$. This work aims to provide more experimental constraints on these scenarios.

\section{Experiment and data analysis}
\label{sec:data_analysis}

The data sample used in this work was collected in LHC Run~1 (2011) and Run~2 (2015), where two beams of Pb nuclei were brought to collide at the centre-of-mass energy of $\sqrt{s_{\rm NN}}=2.76$~TeV and $\sqrt{s_{\rm NN}}=5.02$~TeV, respectively. Products of the collisions were measured by the ALICE detector~\cite{Aamodt:2008zz}. The performance of ALICE is described in Ref.~\cite{Abelev:2014ffa}.

In this analysis the minimum-bias (MB) trigger was used. It is based on the V0 detector consisting of two arrays of 32 scintillator counters, which are installed on each side of the interaction point and cover pseudorapidity\footnote{Pseudorapidity is defined as $\eta=-\ln \left ( \tan  \left (\theta/2 \right ) \right )$, where $\theta$ is the polar angle.} ranges $2.8<\eta<5.1$ (V0A) and $-3.7<\eta<-1.7$ (V0C). The MB trigger required a signal in both V0 detectors within a time window that is consistent with the collision occurring at the centre of the ALICE detector. The event centrality was determined by analysing the amplitude of the signal from the V0 detector with the procedure described in details in Ref.~\cite{Aamodt:2010cz}.

The position of the collision vertex was reconstructed using the signal from the Inner Tracking System (ITS)~\cite{Aamodt:2008zz}. The ITS is composed of six cylindrical layers of silicon detectors and covers $|\eta|<0.9$. Its information can be used for tracking and primary vertex determination. However, in this analysis it was used only for the latter. The primary vertex for an event was required to be within $\pm 8$~cm from the centre of the detector.

The analysis was performed in six centrality~\cite{Abelev:2013qoq} ranges for both collision energies. They are listed in Tab.~\ref{tab:centrality} together with their corresponding average charged-particle multiplicity densities at mid-rapidity $\langle \mathrm{d}N_{\rm ch}/\mathrm{d}\eta \rangle$~\cite{Aamodt:2010cz,Adam:2015ptt}.
\newpage

\begin{center}
\captionof{table}{Centrality ranges and corresponding average charged-particle multiplicity densities at mid-rapidity $\langle \mathrm{d}N_{\rm ch}/\mathrm{d}\eta \rangle$ for Pb--Pb collisions at $\sqrt{s_{\rm NN}}=2.76$~TeV~\cite{Aamodt:2010cz} and $\sqrt{s_{\rm NN}}=5.02$~TeV~\cite{Adam:2015ptt}.}
 \begin{tabular}{ | c | c | c | }
    \hline
    Centrality & $\langle \mathrm{d}N_{\rm ch}/\mathrm{d}\eta \rangle$ $\sqrt{s_{\rm NN}}=2.76$~TeV & $\langle \mathrm{d}N_{\rm ch}/\mathrm{d}\eta \rangle$ $\sqrt{s_{\rm NN}}=5.02$~TeV\\ \hline
    \hline
    0--5\%     & $1601 \pm 60$ & $1943 \pm 53$ \\ \hline
    5--10\%   & $1294 \pm 49$ & $1586 \pm 46$ \\ \hline
    10--20\% & $966 \pm 37$   & $1180 \pm 31$ \\ \hline
    20--30\% & $649 \pm 23$   & $786 \pm 20$ \\ \hline 
    30--40\% & $426 \pm 15$   & $512 \pm 15$ \\ \hline
    40--50\% & $261 \pm 9$   & $318 \pm 12$ \\ \hline
\end{tabular}
\label{tab:centrality}
\end{center}

Charged-particle trajectory (track) reconstruction for both collision energies was performed using the Time Projection Chamber (TPC) detector~\cite{Alme:2010ke}. The TPC is divided by the central electrode into two halves. Each half is capped with a readout plane which is composed of 18 sectors (covering the full azimuthal angle $\varphi$) with 159 padrows placed radially in each sector. A track signal in the TPC consists of space points (clusters), and each of them is reconstructed in one of the padrows. A track was required to be composed of at least 80 clusters to minimise the possibility that a signal left by a single particle is reconstructed as two tracks. The parameters of the track are determined by performing a Kalman fit to a set of clusters. The quality of the fit is determined by calculating the $\chi^2$ which was required to be lower than 4 for every cluster (each cluster has two d.o.f.), in order to select only well fitted tracks.

The identification of primary protons (antiprotons) was performed using the combined information from both the TPC and the Time-OF-Flight (TOF) detectors (a signal from both detectors was required), while the identification of $\Lambda$ ($\overline{\Lambda}$) decay products (charged secondary pions and (anti)protons) required information only from the TPC. TOF is a cylindrical detector composed of Multigap Resistive Proportional Chambers (MRPC) located at $r\cong 380$~cm from the beam axis. Tracks are propagated from the TPC to the TOF and matched to hits in this detector. In the case of both TPC and TOF, the signals (energy loss $\mathrm{d}E/\mathrm{d}x$ for the TPC and the time of flight for the TOF) were compared to the expected ones for a given particle. The measured--expected signal deviation was divided by the appropriate detector resolution $\sigma$. The track was accepted as a proton (pion) if it fell within $3\sigma$ of combined TPC and TOF expected signals for a proton (pion) in a given detector.

Tracks were accepted for analysis if their pseudorapidity range was within the range $|\eta|<0.8$ to avoid regions of the  detector with limited acceptance. The particle identification quality depends on the transverse momentum $p_{\mathrm{T}}$, thus a $p_{\rm T} \in \left [ 0.7,\ 4.0 \right ]$~GeV/$c$ range was used for primary (anti)protons to assure good purity of the sample. To make sure that the sample is not significantly contaminated by secondary particles coming from weak decays and particle--detector interactions, a selection criterion on the Distance of Closest Approach (DCA) to the primary vertex was also applied, separately in the transverse plane ($\rm DCA_{xy}<2.4$~cm) and along the beam axis ($\rm DCA_{z}<3.2$~cm). These criteria were optimised in order to select a high purity sample of (anti)protons. The $p_{\mathrm{T}}$-integrated purity, based on Monte Carlo simulations, of the p ($\rm \overline p$) sample was 95.4\% (95.2\%).

\begin{figure}[!ht]
\centering
\includegraphics[width=0.9\textwidth]{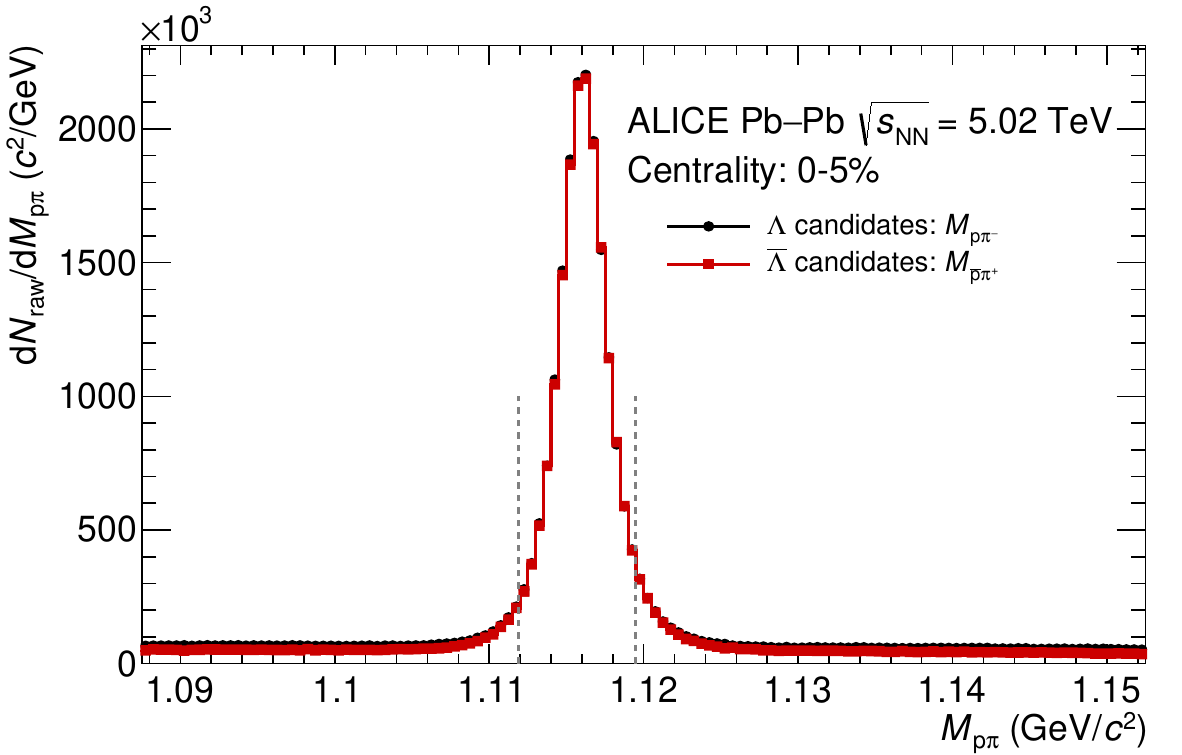}
\caption{Raw invariant mass distribution of $\rm p\uppi^{-}$ ($\rm \overline{p}\uppi^{+}$) pairs used to obtain the $\Lambda$ ($\overline{\Lambda}$) candidates for Pb--Pb collisions at $\sqrt{s_{\rm NN}}=5.02$~TeV in the 0--5\% centrality range. The dashed lines represent the selection width used in the analysis. Note that the mean value of the distribution is slightly shifted from the rest mass of $\Lambda$ ($\overline{\Lambda}$) by $\sim 1$~MeV/$c^2$ due to imperfections
in the energy loss corrections that are applied in the track reconstruction.}
\label{fig:invariant_mass}
\end{figure}

The selection of $\Lambda$ ($\overline{\Lambda}$) is based on their distinctive decay topology in the decay channel $\Lambda\ (\overline{\Lambda})\rightarrow\mathrm{p}\uppi^{-}\ (\mathrm{\overline{p}}\uppi^{+})$, with a branching ratio of 63.9\%~\cite{Patrignani:2016xqp}. The reconstruction process, described in Ref.~\cite{Alessandro:2006yt}, is based on finding candidates made of two secondary tracks having opposite charge and large impact parameter with respect to the interaction point. The purity of $\Lambda$ and $\overline\Lambda$ samples is larger than 95\% within the selected invariant mass range of $|M_{\rm p\uppi}-M_{\Lambda_{\rm PDG}}|\leq 0.0038$~GeV/$c^2$. The $p_{\rm T}$-integrated invariant mass distribution of $\Lambda$ ($\overline{\Lambda}$) candidates is shown in Fig.~\ref{fig:invariant_mass}.

The femtoscopic correlation is measured as a function of the reduced momentum difference of the pair $\vec{k^*} = \frac{1}{2}\left ( \vec{p_1}^* - \vec{p_2}^* \right )$, where $\vec{p_1}^*$ and $\vec{p_2}^*$ denote momenta of the two particles in the pair rest frame. It is defined as

\begin{equation}
C (\vec{k}^* ) = \mathcal{N}\frac {A ( \vec{k}^* )} {B (\vec{k}^* )}.
\label{eq:cfdef2}
\end{equation}

The distribution $A$, called the ``signal'', is constructed from pairs of particles from the same event. The background distribution $B$ is constructed from uncorrelated particles measured with the same single-particle acceptance. In this analysis it was built using the event mixing method with the two particles coming from two different events for which the vertex positions in the beam direction agree within $2$~cm and the multiplicities differ by no more than $1/4$ of the width of the given centrality class for which the correlation function is calculated. Each particle was correlated with particles from 10 other events. The parameter $\mathcal{N}$ is a normalisation factor.

In this work, the analysis is further simplified by performing all measurements as a function of the magnitude of the relative momentum $k^* =  |\vec{k}^*  |$ only. The $\mathcal{N}$ parameter was calculated during the background subtraction procedure described in Sec.~\ref{sec:fitting_procedure}, in a way that the correlation function approaches unity in $k^*\in[0.13, ~1.5]$\,GeV/$c$ for $\rm p\overline{p}$ pairs and in $k^*\in[0.23, ~1.5]$\,GeV/$c$ for $\rm p\overline{\Lambda}$, $\rm \overline{p}\Lambda$, and $\Lambda\overline{\Lambda}$ pairs.
\section{Fitting procedure}
\label{sec:fitting_procedure}

The extraction of the scattering parameters from the measured correlation functions requires a dedicated fitting procedure, which takes into account the strong and Coulomb interaction, depending on a given pair. The fitting formula is chosen appropriately for each baryon--antibaryon pair. Afterwards, a simultaneous fit to all measured pairs, taking into account residual correlations, is performed. The details of the procedure are described below.

The two-particle correlation function in the pair rest frame is defined as~\cite{Koonin:1977fh,Pratt:1990zq}

\begin{equation}
C(\vec{k}^*) = \int S (\vec{r}^* )~\left | {\Psi  ( \vec{k}^{*},\vec{r}^{*} )} \right |^2 \mathrm{d}^3{r}^*,
\label{eq:cf_def}
\end{equation}

\noindent where $S (\vec{r}^*  )$ is the source emission function, $\Psi (\vec{k}^{*},\vec{r}^{*}  )$ is the pair wave function, and $\vec{r}^*$ is the relative separation vector. The source is assumed to have a spherically-symmetric Gaussian distribution according to measurements~\cite{Lisa:2005dd,Aamodt:2011mr}. The pair wave function depends on the interactions between baryons and antibaryons. When only the strong interaction is present, the correlation function can be expressed analytically as a function of the scattering amplitude $f(k^{*})=\left [ \frac{1}{f_0}+\frac{1}{2}d_0k^{*2}-{\rm i}k^{*} \right ]^{-1}$, and the one-dimensional source size $R$. This description is called the Lednick\'y--Lyuboshitz analytical model~\cite{Lednicky:1981su} (see Appendix~\ref{sec:lednicky} for details). In this work, the spin-averaged scattering parameters are obtained, i.e. $\Re f_0$ the real and $\Im f_0$ imaginary parts of the spin-averaged scattering length, and $d_0$ for the real part of the spin-averaged effective range of the interaction. The usual femtoscopic sign convention is used, where a positive $\Re f_0$ corresponds to attractive strong interaction.

\begin{figure}[!ht]
	\centering\includegraphics[width=.9\linewidth]{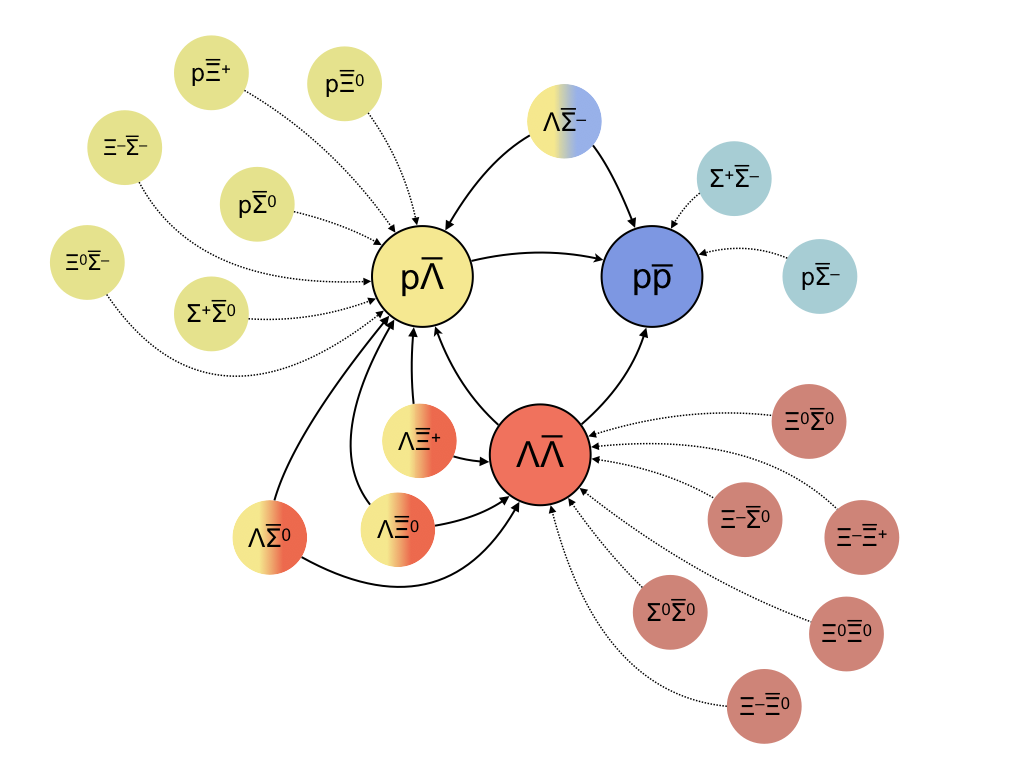}
	\caption{Illustration of the links between different baryon--antibaryon pairs through the residual correlation. Main contributions to the $\rm p\overline{p}$ correlation function are marked in blue, to the $\rm p\overline{\Lambda}$ in yellow and to the $\Lambda\overline{\Lambda}$ in red. Solid lines show connections between studied pairs, while dashed lines present other major residual contributions that are unique for a given system.}
	\label{fig:residual_scheme}
\end{figure}

Accounting for residual correlations is an important ingredient of every correlation function analysis involving baryons. A fraction of observed (anti)baryons comes from decays of heavier (anti)baryons. This is illustrated in Fig.~\ref{fig:residual_scheme}, where the main contributions to the $\rm p\overline{p}$ correlation function are marked in blue, to the $\rm p\overline{\Lambda}$ in yellow and to the $\Lambda\overline{\Lambda}$ in red. In such a case, the correlation function is built for the daughter particles, while the interaction has taken place for the parent baryons. To account for this effect, the fitting formula used in this work contains a sum of correlation functions for each possible combination of (anti)baryons, weighted by the fraction $\lambda$ of given residual pairs. One needs to transform the theoretical correlation function of a pair into the momentum frame of the particles registered in the detector~\cite{Kisiel:2014mma}.

The procedure for the correlation function analysis taking into account residual correlations has been performed before and is described in detail in Ref.~\cite{Adam:2015vja}. The same procedure was carried out in this analysis.

The fractions of residual pairs $\lambda$ were calculated based on the AMPT model~\cite{Lin:2004en} (as well as HIJING~\cite{Wang:1991hta} for evaluation of systematic uncertainties) after full detector simulation, estimating how many reconstructed pairs come from primary particles and what is the percentage of those coming from the given decay. They also take into account other impurities resulting from misidentification or detector effects; therefore, their sum is not equal to unity. The obtained values of fractions are listed in Tab.~\ref{tab:fractions_ampt}. The momentum transformation matrices~\cite{Kisiel:2014mma} were generated using the THERMINATOR 2 model~\cite{Chojnacki:2011hb} for all residual components of all analysed systems. The final correlation function for a $xy$ pair is defined as

\begin{equation}
C_{xy}(k^*) = 1+\sum_i \lambda_i \left[ C_i(k^*)-1 \right],
\end{equation}

\noindent where the sum is over all residual components of the $xy$ pair and $\lambda_i$ and $C_i(k^*)$ are the fraction and the correlation function of $i$-th pair, respectively~\cite{Kisiel:2014mma}.

\begin{table}
\begin{center}
\caption{Fractions of residual components of $\rm p\overline{p}$, $\rm p\overline{\Lambda} \oplus \overline{p}\Lambda$, and $\Lambda\overline{\Lambda}$ correlation functions from Monte Carlo events simulated with AMPT model after full detector simulation. The values in parentheses represent fractions obtained with the HIJING model used for evaluation of systematic uncertainties. Fractions are the same for corresponding antipairs.
}
\begin{tabularx}{\textwidth}{l|X || l|X || l|X}
\hline  \hline
\multicolumn{2}{c||}{$\rm p\overline{p}$} & 
\multicolumn{2}{c||}{$\rm p\overline{\Lambda} \oplus \overline{p}\Lambda$} & 
\multicolumn{2}{c}{$\Lambda\overline{\Lambda}$} \\
Pair & $\lambda$ &
Pair & $\lambda$ &
Pair & $\lambda$ \\
\hline 
$\rm p\overline{p}$             & 0.25 (0.32) &
$\rm p\overline{\Lambda}$       & 0.29 (0.28) &
$\Lambda\overline{\Lambda}$     & 0.37 (0.24)\\

$\rm p\overline{\Lambda}$      & 0.12 (0.19) &
$\Lambda\overline{\Lambda}$    & 0.08 (0.09)&
$\Lambda\overline{\Xi}^+$      & 0.04 (0.06)\\

$\rm p\overline{\Sigma}^-$         & 0.04 (0.04)&
$\Lambda\overline{\Sigma}^-$   & 0.03 (0.02)&
$\Lambda\overline{\Xi}^0$      & 0.03 (0.05)\\
 
$\Lambda\overline{\Lambda}$    & 0.02 (0.03)&
$\rm p\overline{\Xi}^{0/+}$        & 0.02 (0.03)&
$\Lambda\overline{\Sigma}^0$   & $<0.01$ (0.20)\\

$\Lambda\overline{\Sigma}^-$   & 0.01 (0.01)&
$\rm p\overline{\Sigma}^0$          & $<0.01$ (0.12)&
$\Sigma^0\overline{\Sigma}^0$   & $<0.01$ (0.05)\\
  
$\Sigma^+\overline{\Sigma}^-$  & $<0.01$ ($<0.01$)& 
$\Lambda\overline{\Sigma}^0$    & $<0.01$ (0.04)&
$\Xi^{0/-}\overline{\Sigma}^0$  & $<0.01$ (0.02)\\         
&&
$\Lambda\overline{\Xi}^{0/+}$   & $<0.01$ (0.01)&
 $\Xi^{0/-}\overline{\Xi}^{0/+}$ & $<0.01$ ($<0.01$)\\
&&
$\Sigma^+\overline{\Sigma}^0$   & $<0.01$ ($<0.01$)&
& \\
&&
$\Xi^{0/-}\overline{\Sigma}^+$  & $<0.01$ ($<0.01$)&
& \\
\hline \hline
\end{tabularx}
\addtocounter{table}{-1}
\label{tab:fractions_ampt}
\end{center}
\end{table}

Correlation functions were obtained for four baryon--antibaryon pair systems $\rm p\overline{p}$, \mbox{$\rm p\overline{\Lambda}$}, $\rm\overline{p}\Lambda$, and $\Lambda\overline{\Lambda}$. Since the correlation functions $\rm p\overline{\Lambda}$ and $\rm\overline{p}\Lambda$ were found to be consistent with each other within the uncertainties, they were always combined and are further denoted as $\rm p\overline{\Lambda} \oplus \overline{p}\Lambda$. A simultaneous fit is desirable because of the presence of residual correlations which link different pairs. Three sets of unknown scattering parameters, components of the analytical formula (\ref{eq:lednicky}) used in the fit, were introduced for $\rm p\overline{\Lambda} \oplus \overline{p}\Lambda$, $\rm \Lambda\overline{\Lambda}$ as well as heavier, not measured directly, baryon--antibaryon pairs, further referred to as $\rm B\overline{B}$, consisting of pairs containing $\Sigma$ and $\Xi$ baryons. The $\rm p\overline{p}$ system was used as a reference. However, due to the presence of the Coulomb interaction and coupled channels, the analytical description is no longer valid in this case (see Appendix~\ref{sec:lednicky} for details). Nevertheless, coupled-channel effects become negligible for large sources as the ones obtained in Pb-Pb collision systems. The theoretical $\rm p\overline{p}$ correlation functions were obtained by generating $\rm p\overline{p}$ pairs with the THERMINATOR 2 model and by applying weights accounting for the final state interactions with an approximate treatment of the $\rm n\overline{n}$ coupled channel, using a numerical model by R. Lednick\'y~\cite{Lednicky:1981su,Lednicky:2005af} with experimental constraints on strong interaction parameters from previous measurements~\cite{Batty:1989gg,Kerbikov:1992xt,Pignone:1994uj}.

The source sizes for primary $\rm p\overline{p}$, $\rm p\overline{\Lambda}\oplus\overline{p}\Lambda$, and $\Lambda\overline{\Lambda}$ pairs were taken from previous measurements of other baryon--baryon and meson--meson pairs~\cite{Adam:2015vja}. We assume that the one-dimensional source size $R$, for each pair, depends on the transverse mass of the pair, $m_{\rm T}=\sqrt{m^2+p_{\rm T}^2}$, and on the charged-particle multiplicity $N_{\rm ch}$ ~\cite{Adam:2015vna} following the relations
\begin{equation}
R(m_{\rm T};N_{\rm ch}) = a(N_{\rm ch}) m_{\rm T}^{\gamma}, 
\label{eq:m_Tscale}
\end{equation}
 and 
\begin{equation}
 R(N_{\rm ch};m_{\rm T}) = \alpha(m_{\rm T})\sqrt[3]{N_{\rm ch}}+\beta(m_{\rm T}),
 \label{eq:nchscale}
  \end{equation}
 where the $\gamma$ exponent and the $a(N_{\rm ch}), \alpha(m_{\rm T})$ and $\beta(m_{\rm T})$ functions are empirical and include the constraint of the minimum possible source size ($N_{\rm ch}=1$) being equal to the proton radius, $R_{\rm p} \approx 0.88$~fm~\cite{Patrignani:2016xqp}. The relations~(\ref{eq:m_Tscale}) and~(\ref{eq:nchscale}) are used for all pairs, including those contributing via weak decays.

The experimental correlation function is also affected by phenomena other than the strong and Coulomb interactions, such as jets and elliptic flow~\cite{Kisiel:2017gip,Aamodt:2011kd,Graczykowski:2014tsa}. Those effects are treated as a background. For each experimental function, a background fit was performed in a $k^*$ region where femtoscopic effects are not prominent. It was found, using the THERMINATOR 2 model, that the results are not dependent on the $k^*$ fit range when the background is fitted by a third order polynomial. Next, the estimated background was subtracted from the experimental correlation function. The procedure flattens the function for higher $k^*$ and the slope is larger for less central collisions, which is consistent with elliptic flow, as it should be more prominent for semi-central collisions and less for central collisions~\cite{Kisiel:2017gip}. 

\begin{figure}
\centering\includegraphics[trim={0cm 0 8.9cm 0},clip, width=.49\linewidth]{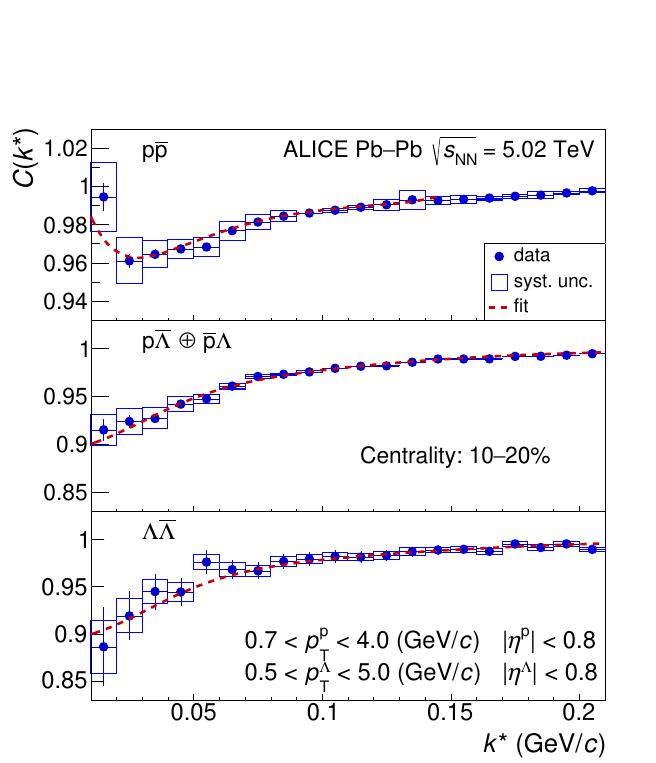}
\centering\includegraphics[trim={8.9cm 0 0cm 0},clip, width=.49\linewidth]{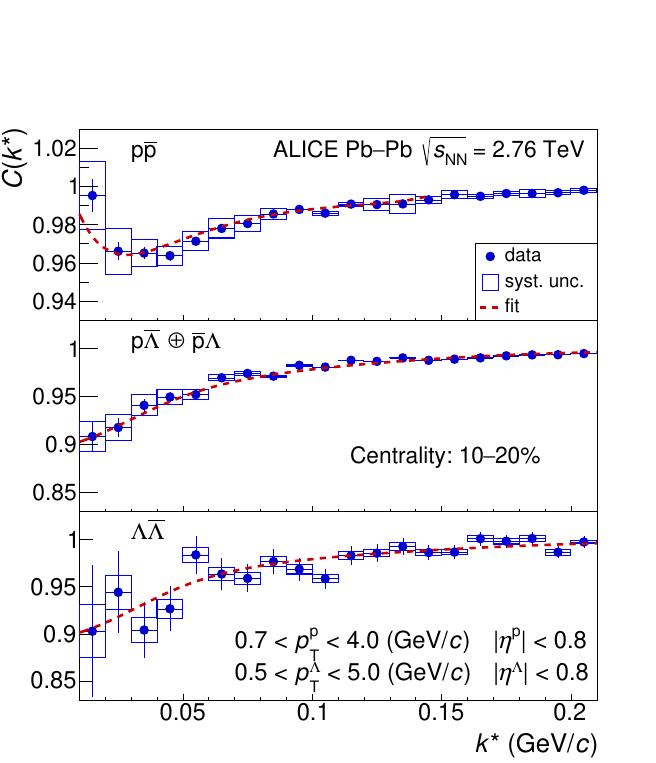}

\caption{Correlation functions of $\rm p\overline{p}$, $\rm p\overline{\Lambda}\oplus\overline{p}\Lambda$, and $\Lambda\overline{\Lambda}$ pairs for Pb--Pb collisions at $\sqrt{s_{\rm NN}}=5.02$~TeV  (left) and $\sqrt{s_{\rm NN}}=2.76$~TeV (right) together with the simultaneous femtoscopic fit for the 10--20\% centrality class.}
\label{fig:data_fit_bab}
\end{figure}

As an example, the correlation functions for $\rm p\overline{p}$, $\rm p\overline{\Lambda}\oplus\overline{p}\Lambda$ and $\Lambda\overline{\Lambda}$ pairs for the 10--20\% centrality interval and two collision energies are represented together with the simultaneous fit in Fig.~\ref{fig:data_fit_bab}.

The momentum resolution effect was investigated with Monte Carlo simulations by creating a two-dimensional matrix of generated and reconstructed $k^{*}$. Each slice of the distribution was then fitted with a Gaussian function. Within the $k^{*}$ region of interest the width of the Gaussian function is constant; therefore, the fitting function was smeared with a Gaussian with a width constant in $k^{*}$.
\section{Results}
\label{sec:results}
\begin{figure}[!ht]
\centering
\includegraphics[width=0.9\textwidth]{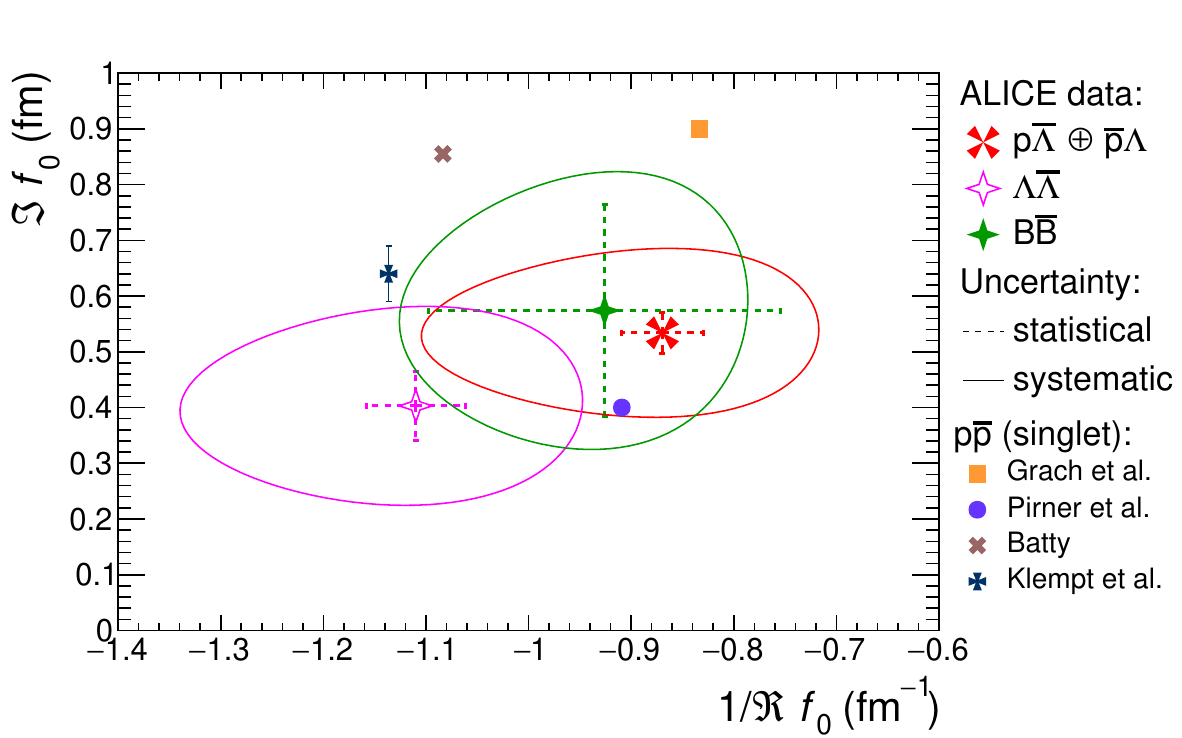}
\includegraphics[width=0.9\textwidth]{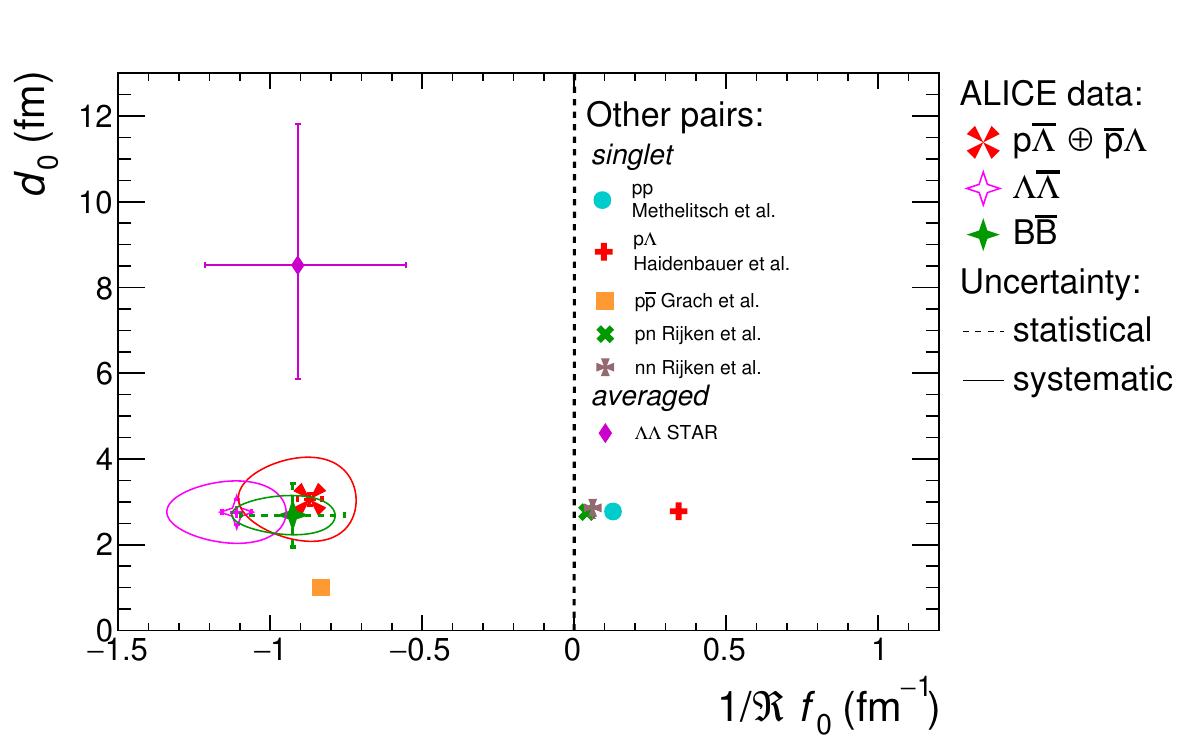}
\caption[Caption for LOF]{(Top) Comparison of extracted spin-averaged scattering parameters $\Re f_{0}$ and $\Im f_{0}$ for $\rm p\overline{\Lambda}\oplus \overline{p}\Lambda$, $\Lambda\overline{\Lambda}$ pairs and for effective $\rm B\overline{B}$ pairs, with previous analyses of $\rm p\overline{p}$ pairs (singlet)~\cite{Grach:1987rp,Pirner:1991mn,Klempt:2002ap,Batty:1989gg}. (Bottom) Comparison of extracted spin-averaged scattering parameters $\Re f_{0}$ and $d_{0}$ for $\rm p\overline{\Lambda}\oplus \overline{p}\Lambda$, $\Lambda\overline{\Lambda}$ pairs as well as effective $\rm B\overline{B}$, with selected previous analyses of other pairs: $\rm pp$ (singlet)~\cite{Mathelitsch:1984hq}, $\rm p\overline{p}$ (singlet)~\cite{Grach:1987rp}, $\rm pn$ (singlet)~\cite{Rijken:2010zzb}, $\rm nn$ (singlet)~\cite{Rijken:2010zzb}, $\rm p\Lambda$ (singlet)~\cite{Haidenbauer:2013oca}, and $\Lambda\Lambda$ (spin-averaged)~\cite{Adamczyk:2014vca}. (Note that the measurement of the $\Lambda\Lambda$ scattering parameters by the STAR experiment~\cite{Adamczyk:2014vca} did not account for residual correlations. The recent analysis of $\Lambda\Lambda$ correlations by the ALICE Collaboration~\cite{Acharya:2018gyz}, properly taking into account those correlations, disfavours the STAR results.)}
\label{fig:params_bab_comparison}
\end{figure}

The strong-interaction scattering parameters $\Re f_0$, $\Im f_0$, and $d_0$ for $\rm p\overline{p}$, $\rm \overline{p}\Lambda \oplus p\overline{\Lambda}$,  $\Lambda\overline{\Lambda}$, and $\rm B\overline{B}$ pairs resulting from the simultaneous fit are summarised in Tab.~\ref{tab:multi_fit_summary_2015}
and plotted in Fig.~\ref{fig:params_bab_comparison} together with statistical (bars) and systematic (ellipses) uncertainties\footnote{Details of the systematic uncertainty estimation are discussed in Appendix~\ref{sec:systematic_uncertainty}.}. Figure~\ref{fig:params_bab_comparison} also shows  scattering parameters for various baryon--baryon and baryon--antibaryon pairs extracted in previous studies~\cite{Kerbikov:2004gs,Haidenbauer:2013oca,Adamczyk:2014vca,Czerwinski:2014yot}.

As the simultaneous fit yields similar values, within uncertainties, of parameters for $\rm \overline{p}\Lambda \oplus p\overline{\Lambda}$ and $\Lambda\overline{\Lambda}$, as well as heavier $\rm B\overline{B}$ pairs, one can perform a fit assuming a single set of parameters for all those systems. By doing so there is practically no change in the results; in particular, the reduced $\chi^2\approx 1.83$ ($p<0.00001$) of the first fit becomes $\chi^2\approx 1.87$ and other scattering parameters change very slightly, within systematic uncertainties. This test confirms that the data points can be correctly described when one assumes that all baryon--antibaryon pairs have similar values of the scattering length and the effective range of the strong interaction.

\begin{center}
\begin{table}[!th]
\flushleft
\normalsize
\centering
\caption{Values of the spin-averaged scattering parameters $\Re f_{0}$, $\Im f_{0}$, and $d_{0}$ for $\rm p\overline{\Lambda} \oplus \overline{p}\Lambda$ and $\Lambda\overline{\Lambda}$ pairs, as well as effective parameters accounting for heavier baryon--antibaryon ($\rm B\overline{B}$) pairs not measured directly, extracted from the simultaneous fit.
}
\begin{tabular}[t]{ | c | c | c | c |} \hline
Parameter  & $\rm \overline{p}\Lambda \oplus p\overline{\Lambda}$ & $\Lambda\overline{\Lambda}$  & $\rm B\overline{B}$      \\ \hline\hline
$\Re{f_0}$~(fm) & $-1.15_{\pm 0.05\ ({\rm stat.})}^{\pm 0.23\ ({\rm syst.})}$     & $-0.90_{\pm 0.04\ ({\rm stat.})}^{\pm 0.16\  ({\rm syst.})}$  &$-1.08_{\pm 0.20\ ({\rm stat.})}^{\pm 0.11\ ({\rm syst.})}$ \\ \hline
$\Im{f_0}$~(fm) & $0.53_{\pm\ 0.04\ ({\rm stat.})}^{\pm\ 0.15\ ({\rm syst.})}$      & $0.40_{\pm 0.06\ ({\rm stat.})}^{\pm 0.18\ ({\rm syst.})}$ & $0.57_{\pm 0.19\ (\rm stat)}^{\pm 0.25\ ({\rm syst.})}$  \\ \hline
$d_0$~(fm) & $3.06_{\pm 0.14\ ({\rm stat.})}^{\pm 0.98\ ({\rm syst.})}$      & $2.76_{\pm 0.29\ ({\rm stat.})}^{\pm 0.73\ ({\rm syst.})}$  & $2.69_{\pm 0.74\ ({\rm stat.})}^{\pm 0.46\ ({\rm syst.})}$ \\ \hline
\end{tabular}
\label{tab:multi_fit_summary_2015}
\end{table}
\end{center}


\section{Discussion}
\label{sec:conclusions}

Femtoscopic correlation functions for $\rm p\overline{p}$, $\rm p\overline{\Lambda}\oplus\overline{p}\Lambda$ and $\Lambda\overline{\Lambda}$ have been measured in Pb--Pb collisions at energies of $\sqrt{s_{\rm NN}}=2.76$~TeV and $\sqrt{s_{\rm NN}}=5.02$~TeV registered by the ALICE experiment. The analysis was performed in six centrality intervals, yielding 36 correlation functions in total.

For the first time parameters of the strong interaction, the scattering length and the effective range, were extracted for $\rm p\overline{\Lambda}\oplus\overline{p}\Lambda$ and $\Lambda\overline{\Lambda}$ pairs. Moreover, parameters for heavier baryon--antibaryon pairs, which were not measured directly, were estimated. 

Several conclusions can be drawn from the extracted parameters. The real and imaginary parts of the scattering length, $\Re f_0$ and $\Im f_0$, and the effective interaction range, $d_0$, have similar values for all baryon--antibaryon pairs at low $k^{*}$. Therefore, the data can be described using the same parameters for all studied pairs, which provides a valuable input for theoretical heavy-ion collisions models. Note that the assumption used in the UrQMD model, namely that $\Im f_0$ is the same for different baryon--antibaryon pairs as a function of the centre-of-mass energy of the pair, means that the inelastic cross section would be different at the same relative pair momentum $k^{*}$.

A significant non-zero imaginary part of the scattering length $\Im f_0$ indicates the presence of the inelastic channel of the interaction, which in the case of baryon--antibaryon includes the annihilation process.

The negative value of the real part of the scattering length, $\Re f_0$, obtained for all baryon--antibaryon pairs may have one of a two meanings: either the strong interaction is repulsive, or a bound state can be formed. The significant magnitude of the imaginary part of the scattering length, $\Im f_0$, shows that baryon--antibaryon scattering may occur through inelastic processes (annihilation). In the UrQMD model, three scenarios can be considered~\cite{Kisiel:2014mma}: i) all baryon--antibaryon pairs annihilate similarly at the same relative momentum $k^{*}$; ii) $\Im f_0$ is the same for all baryon--antibaryon pairs, but expressed as a function of the pair centre-of-mass energy, meaning that $\Im f_0$ is smaller for baryon--antibaryon pairs of higher total pair mass; iii) the inelastic cross section is increased for every matching quark--antiquark pair in the baryon--antibaryon system. In this scenario, in the specific case of this work, $\Im f_0$ for $\rm p\overline{\Lambda}\oplus\overline{p}\Lambda$ should be lower than for $\rm p\overline{p}$ and $\Lambda\overline{\Lambda}$, which is not observed. UrQMD by default uses scenario ii) to model the baryon--antibaryon annihilation, which in our case would lead to a decrease of $\Im f_0$ while going from $\rm p\overline{p}$ to $\Lambda\overline{\Lambda}$ pairs; however, similar values of $\Im f_0$ for all baryon--antibaryon pairs reported in this work favour scenario i).

Inelastic scattering is compatible with a bound state, where the baryon and antibaryon create a short-lived resonance which decays strongly into three mesons. Evidence for a process in which a particle in the mass range of 2150--2260~MeV/$c^2$ decays into a kaon and two pions has been reported by various experiments in the past and listed by the Particle Data Group (PDG) as $\rm K_2(2250)$~\cite{Patrignani:2016xqp}. The reported mass is slightly above the $\rm p\overline{\Lambda}$ threshold, the width of the resonance is compatible with a strongly decaying system and the decay products match the valence quark content of the $\rm p\overline{\Lambda}$ pair. A~nucleon--antihyperon system has also been listed by PDG as $\rm K_3(2320)$, with proton and $\overline{\Lambda}$ in the final state, which corresponds to a bound state undergoing an elastic scattering. The results presented in this paper support the existence of baryon--antibaryon bound states such as $\rm K_2(2250)$ and $\rm K_3(2320)$. Further studies can provide more evidence on the existence of those states.

Finally, negative values of the extracted real part of the scattering length $\Re f_0$ show either that the interaction between baryons and antibaryons is repulsive, or that baryon--antibaryon bound states can be formed. Combined with the non-zero imaginary part $\Im f_0$ which, as mentioned earlier, is associated with the inelastic processes, it favours the bound states scenario over the repulsive interaction. In that case a baryon--antibaryon pair would form a resonance decaying into a group of particles different from the original ones (for instance, $\mathrm{p}\overline{\Lambda} \rightarrow \rm X \rightarrow \rm K^+\uppi^+\uppi^-$, where $\rm X$ is the hypothetical baryon--antibaryon bound state). Further studies will shed more light on existence of such particles. The scenario of a repulsive interaction is not completely ruled out, but it would manifest in experiments as a systematic spatial separation of matter and antimatter, never observed before. 

In summary, the strong-interaction cross section parameters (the scattering length and the effective range) of strange baryon--antibaryon pairs have been measured at low relative pair momentum using the femtoscopic technique. They were found to be the same within the systematic uncertainties for all studied pairs and compatible with the $\rm p\overline{p}$ parameters measured in other experiments. Therefore, a global picture of the baryon--antibaryon annihilation proceeding in a very similar way, regardless of the strange-quark content, is suggested. Finally, the results are consistent with the formation of baryon--antibaryon bound states. Future searches for such particles will therefore be of crucial importance.

\newenvironment{acknowledgement}{\relax}{\relax}
\begin{acknowledgement}
\section*{Acknowledgements}

The ALICE Collaboration would like to thank all its engineers and technicians for their invaluable contributions to the construction of the experiment and the CERN accelerator teams for the outstanding performance of the LHC complex.
The ALICE Collaboration gratefully acknowledges the resources and support provided by all Grid centres and the Worldwide LHC Computing Grid (WLCG) collaboration.
The ALICE Collaboration acknowledges the following funding agencies for their support in building and running the ALICE detector:
A. I. Alikhanyan National Science Laboratory (Yerevan Physics Institute) Foundation (ANSL), State Committee of Science and World Federation of Scientists (WFS), Armenia;
Austrian Academy of Sciences, Austrian Science Fund (FWF): [M 2467-N36] and Nationalstiftung f\"{u}r Forschung, Technologie und Entwicklung, Austria;
Ministry of Communications and High Technologies, National Nuclear Research Center, Azerbaijan;
Conselho Nacional de Desenvolvimento Cient\'{\i}fico e Tecnol\'{o}gico (CNPq), Universidade Federal do Rio Grande do Sul (UFRGS), Financiadora de Estudos e Projetos (Finep) and Funda\c{c}\~{a}o de Amparo \`{a} Pesquisa do Estado de S\~{a}o Paulo (FAPESP), Brazil;
Ministry of Science \& Technology of China (MSTC), National Natural Science Foundation of China (NSFC) and Ministry of Education of China (MOEC) , China;
Croatian Science Foundation and Ministry of Science and Education, Croatia;
Centro de Aplicaciones Tecnol\'{o}gicas y Desarrollo Nuclear (CEADEN), Cubaenerg\'{\i}a, Cuba;
Ministry of Education, Youth and Sports of the Czech Republic, Czech Republic;
The Danish Council for Independent Research | Natural Sciences, the Carlsberg Foundation and Danish National Research Foundation (DNRF), Denmark;
Helsinki Institute of Physics (HIP), Finland;
Commissariat \`{a} l'Energie Atomique (CEA), Institut National de Physique Nucl\'{e}aire et de Physique des Particules (IN2P3) and Centre National de la Recherche Scientifique (CNRS) and Rl\'{e}gion des  Pays de la Loire, France;
Bundesministerium f\"{u}r Bildung, Wissenschaft, Forschung und Technologie (BMBF) and GSI Helmholtzzentrum f\"{u}r Schwerionenforschung GmbH, Germany;
General Secretariat for Research and Technology, Ministry of Education, Research and Religions, Greece;
National Research, Development and Innovation Office, Hungary;
Department of Atomic Energy Government of India (DAE), Department of Science and Technology, Government of India (DST), University Grants Commission, Government of India (UGC) and Council of Scientific and Industrial Research (CSIR), India;
Indonesian Institute of Science, Indonesia;
Centro Fermi - Museo Storico della Fisica e Centro Studi e Ricerche Enrico Fermi and Istituto Nazionale di Fisica Nucleare (INFN), Italy;
Institute for Innovative Science and Technology , Nagasaki Institute of Applied Science (IIST), Japan Society for the Promotion of Science (JSPS) KAKENHI and Japanese Ministry of Education, Culture, Sports, Science and Technology (MEXT), Japan;
Consejo Nacional de Ciencia (CONACYT) y Tecnolog\'{i}a, through Fondo de Cooperaci\'{o}n Internacional en Ciencia y Tecnolog\'{i}a (FONCICYT) and Direcci\'{o}n General de Asuntos del Personal Academico (DGAPA), Mexico;
Nederlandse Organisatie voor Wetenschappelijk Onderzoek (NWO), Netherlands;
The Research Council of Norway, Norway;
Commission on Science and Technology for Sustainable Development in the South (COMSATS), Pakistan;
Pontificia Universidad Cat\'{o}lica del Per\'{u}, Peru;
Ministry of Science and Higher Education and National Science Centre, Poland;
Korea Institute of Science and Technology Information and National Research Foundation of Korea (NRF), Republic of Korea;
Ministry of Education and Scientific Research, Institute of Atomic Physics and Ministry of Research and Innovation and Institute of Atomic Physics, Romania;
Joint Institute for Nuclear Research (JINR), Ministry of Education and Science of the Russian Federation, National Research Centre Kurchatov Institute, Russian Science Foundation and Russian Foundation for Basic Research, Russia;
Ministry of Education, Science, Research and Sport of the Slovak Republic, Slovakia;
National Research Foundation of South Africa, South Africa;
Swedish Research Council (VR) and Knut \& Alice Wallenberg Foundation (KAW), Sweden;
European Organization for Nuclear Research, Switzerland;
National Science and Technology Development Agency (NSDTA), Suranaree University of Technology (SUT) and Office of the Higher Education Commission under NRU project of Thailand, Thailand;
Turkish Atomic Energy Agency (TAEK), Turkey;
National Academy of  Sciences of Ukraine, Ukraine;
Science and Technology Facilities Council (STFC), United Kingdom;
National Science Foundation of the United States of America (NSF) and United States Department of Energy, Office of Nuclear Physics (DOE NP), United States of America.
\end{acknowledgement}

\bibliographystyle{utphys}
\bibliography{bibliography}

\newpage
\appendix
\section{Lednick\'y--Lyuboshitz model}
\label{sec:lednicky}
The wave function of the pair, $\Psi  ( \vec{k}^{*},\vec{r}^{*} )$, in Eq.~(\ref{eq:cf_def}), depends on the two-particle interaction. Baryons interact with anti-baryons via the strong and, if they carry a non-zero electric charge, the Coulomb force. In such a scenario, the interaction of two non-identical particles is given by the Bethe--Salpeter amplitude, corresponding to the solution of the quantum scattering problem taken with the inverse time direction:

\begin{equation}
\Psi^{(+)}_{-{\vec{k}^{*}}}({\vec{{r}^{*}},{\vec{k}^{*}}}) = \sqrt{A_{\rm C} (\eta)}
\frac{1}{\sqrt{2}} \left [ {\rm e}^{-{\rm i} \vec{ k}^{*}\cdot{\vec{r}^{*}}} {\rm F}(-{\rm i} \eta, 1,
  {\rm i} \zeta^{+}) + f_{\rm C}(\vec{k}^*)\frac{\tilde{G}(\rho,\eta)}{{r}^*} \right ],
\label{eq:fullpsi2}
\end{equation}
where $A_{\rm C}$ is the Gamow factor, $\zeta^{\pm} = k^{*} r^{*} (1 \pm \cos{\theta^{*}})$, $\eta = 1/(k^{*} a_{\rm C})$, $\rm F$ is the confluent hypergeometric function, and $\tilde{G}$ is the combination of the regular and singular $\rm S$-wave Coulomb functions. $\theta^{*}$ is the angle between the pair relative momentum and relative position in the pair rest frame, while $a_{\rm C}$ is the Bohr radius of the pair. The component $f_{\rm C}$ is the strong-scattering amplitude, modified by the Coulomb interaction.

When only the strong interaction is present, the correlation function can be expressed analytically as a function of the scattering amplitude $f(k^{*})=\left [ \frac{1}{f_0}+\frac{1}{2}d_0k^{*2}-{\rm i}k^{*} \right ]^{-1}$, and the one-dimensional source size $R$. This description is called the Lednick\'y--Lyuboshitz analytical model~\cite{Lednicky:1981su}:   

\begin{equation}
\label{eq:lednicky}
C(k^*) = 1+ \sum_{\sigma} \rho_{\sigma} \left[ \frac{1}{2} \left | {\frac{f(k^*)}{R}} \right |^2 \left( 1-\frac{d_0^{\sigma}}{2\sqrt{\rm {\uppi}}R} \right) + \frac{2\Re{f(k^*)}}{\sqrt{\rm {\uppi}}R} F_1(2k^*R) - \frac{\Im{f(k^*)}}{R}F_2(2k^*R) \right],
\end{equation}

\noindent where the sum is over all pair-spin configurations ${\sigma}$, with weights $\rho_{\sigma}$ (a real number) being $1/4$ and $3/4$ for singlet and triplet states, respectively, and $F_1(z)=\int_0^z({\rm e}^{x^2-z^2}/z)\mathrm{d}x$ and $F_2(z)=(1-{\rm e}^{-z^2})/z$. When the Coulomb interaction is also present, e.g., in the $\rm p\overline{p}$ case, the source emission function is numerically integrated with the pair wave function containing a modified scattering amplitude~\cite{Lednicky:2005tb}:

\begin{equation}
f_{\rm C}(k^{*})=\left [ \frac{1}{f_0}+\frac{1}{2}d_0k^{*2}-{\rm i}k^{*}-
\frac{2}{a_{\rm C}}h(\eta) - {\rm i}k^*A_{\rm C}(\eta) \right ]^{-1},
\end{equation}
\noindent where
$h(\eta)=\eta^2\sum\limits_{n=1}^{ \infty}[n(n^2+\eta^2)]^{-1}
-\mathrm{{\upgamma}} -\ln|\eta|$
($\mathrm{{\upgamma}} = 0.5772$ is the Euler constant).

The description becomes more complicated when coupled channels (such as $\rm n\overline{n}\rightarrow p\overline{p}$ in the $\rm p\overline{p}$ system) are present. For details see Ref.~\cite{Lednicky:2005tb,Haidenbauer:2018jvl}.
\section{Systematic uncertainties}
\label{sec:systematic_uncertainty}

The analysis was also performed on tracks reconstructed using the information from both the ITS and the TPC, as opposed to using those having the information from the TPC only. The correlation functions obtained from the analysis of those tracks were fitted with the procedure described in Sec.~\ref{sec:fitting_procedure}. Differences on extracted scattering parameters are between 4\% and 17\%, depending on the studied pair and the scattering parameter.

In addition, several components of the fit procedure were varied. Shifting the correlation function normalisation range in $k^{*}$ by $\pm 0.1$~GeV/$c$ yields almost no change on the extracted scattering parameters (maximum 1\%). A change of the background parametrisation from the third to the fourth-order polynomial results in differences of up to 19\% for $\Im f_0$ and below 10\% for other parameters. The second-order polynomial was also tested but it fails to describe the low $k^{*}$ region and therefore cannot be used to extract reliable information. Moreover, the use of residual pair fractions calculated from the HIJING model~\cite{Wang:1991hta} instead of AMPT resulted in changes of up to 19\% for $d_0$, up to 16\% for $\Im f_0$, and below 10\% for $\Re f_0$. Variation of source sizes obtained from transverse-mass and multiplicity scalings by $\pm 5$\% resulted in changes of up to 13\% for $\Re f_0$, up to 36\% for $\Im f_0$, and up to 20\% for $d_0$. Moreover, the width of the Gaussian distribution accounting for momentum resolution was varied by $\pm 30\%$ which results in systematic uncertainty of up to 11\%. 

Contributions to the systematic uncertainty on the extracted scattering parameters are summarised in Tab.~\ref{tab:systerrors}. Since those components are correlated, the total systematic uncertainties are represented as covariance ellipses in the final plots.

\begin{table}[!htb]
\begin{center}
\caption{List of contributions to the systematic uncertainty of the scattering parameters. Values are averaged over collision energies and centrality ranges. }
\label{tab:systerrors}
\begin{tabular}{l|c|c|c}
\hline
\multicolumn{4}{c}{\bf $\mathrm{p}\overline{\Lambda}\oplus\overline{\mathrm{p}}\Lambda$} \\ 
\hline
\hline
{\bf Uncertainty source} & $\Re{f_0}$ (\%) & $\Im{f_0}$ (\%) & $d_0$ (\%) \\
\hline
Normalisation range & $<1$ & $<1$ & $<1$\\
Background parametrisation & $<1$ & $2$ & 3\\
Fit range dependence & 3 & 8 & 14 \\
Fractions of residual pairs & 10 & 8 & 19 \\
Momentum resolution correction & 7 & 11 & 4 \\
Track selection & 11 & 14 & 4 \\
Source size variation & 9 & 18 & 20 \\
\hline
\hline
\multicolumn{4}{c}{\bf $\mathrm{\Lambda}\overline{\Lambda}$} \\ 
\hline
\hline
{\bf Uncertainty source} & $\Re{f_0}$ (\%) & $\Im{f_0}$ (\%) & $d_0$ (\%) \\
\hline
Normalisation range & $<1$ & $<1$ & $<1$\\
Background parametrisation & 6 & 19 & 2\\
Fit range dependence & 2 & 4 & 5 \\
Fractions of residual pairs & 6 & 15 & 18 \\
Momentum resolution correction & 4 & 7 & 2 \\
Track selection & 7 & 17 & 4 \\
Source size variation & 12 & 35 & 19 \\
\hline
\hline
\multicolumn{4}{c}{\bf $\mathrm{B}\overline{\rm B}$} \\ 
\hline
\hline
{\bf Uncertainty source} & $\Re{f_0}$ (\%) & $\Im{f_0}$ (\%) & $d_0$ (\%) \\
\hline
Normalisation range & $<1$ & 1 & 1\\
Background parametrisation & 6 & 17 & 6\\
Fit range dependence & 6 & 12 & 11 \\
Fractions of residual pairs & 7 & 19 & 8 \\
Momentum resolution correction & 3 & 3 & 1 \\
Track selection & 9 & $<1$ & 12 \\
Source size variation & 13 & 36 & 9 \\
\hline
\end{tabular}
\end{center}
\end{table}

\newpage

\section{The ALICE Collaboration}
\label{app:collab}

\begingroup
\small
\begin{flushleft}
S.~Acharya\Irefn{org141}\And 
D.~Adamov\'{a}\Irefn{org93}\And 
S.P.~Adhya\Irefn{org141}\And 
A.~Adler\Irefn{org74}\And 
J.~Adolfsson\Irefn{org80}\And 
M.M.~Aggarwal\Irefn{org98}\And 
G.~Aglieri Rinella\Irefn{org34}\And 
M.~Agnello\Irefn{org31}\And 
N.~Agrawal\Irefn{org10}\And 
Z.~Ahammed\Irefn{org141}
% ...
\And
\renewcommand\labelenumi{\textsuperscript{\theenumi}~}

\section*{Affiliation notes}
\renewcommand\theenumi{\roman{enumi}}
\begin{Authlist}
\item \Adef{org*}Deceased
\item \Adef{orgI}Dipartimento DET del Politecnico di Torino, Turin, Italy
\item \Adef{orgII}M.V. Lomonosov Moscow State University, D.V. Skobeltsyn Institute of Nuclear, Physics, Moscow, Russia
\item \Adef{orgIII}Department of Applied Physics, Aligarh Muslim University, Aligarh, India
\item \Adef{orgIV}Institute of Theoretical Physics, University of Wroclaw, Poland
\end{Authlist}

\section*{Collaboration Institutes}
\renewcommand\theenumi{\arabic{enumi}~}
\begin{Authlist}
\item \Idef{org1}A.I. Alikhanyan National Science Laboratory (Yerevan Physics Institute) Foundation, Yerevan, Armenia
\item \Idef{org2}Bogolyubov Institute for Theoretical Physics, National Academy of Sciences of Ukraine, Kiev, Ukraine
\item \Idef{org3}Bose Institute, Department of Physics  and Centre for Astroparticle Physics and Space Science (CAPSS), Kolkata, India
\item \Idef{org4}Budker Institute for Nuclear Physics, Novosibirsk, Russia
\item \Idef{org5}California Polytechnic State University, San Luis Obispo, California, United States
\item \Idef{org6}Central China Normal University, Wuhan, China
\item \Idef{org7}Centre de Calcul de l'IN2P3, Villeurbanne, Lyon, France
\item \Idef{org8}Centro de Aplicaciones Tecnol\'{o}gicas y Desarrollo Nuclear (CEADEN), Havana, Cuba
\item \Idef{org9}Centro de Investigaci\'{o}n y de Estudios Avanzados (CINVESTAV), Mexico City and M\'{e}rida, Mexico
\item \Idef{org10}Centro Fermi - Museo Storico della Fisica e Centro Studi e Ricerche ``Enrico Fermi', Rome, Italy
\item \Idef{org11}Chicago State University, Chicago, Illinois, United States
\item \Idef{org12}China Institute of Atomic Energy, Beijing, China
\item \Idef{org13}Chonbuk National University, Jeonju, Republic of Korea
\item \Idef{org14}Comenius University Bratislava, Faculty of Mathematics, Physics and Informatics, Bratislava, Slovakia
\item \Idef{org15}COMSATS University Islamabad, Islamabad, Pakistan
\item \Idef{org16}Creighton University, Omaha, Nebraska, United States
\item \Idef{org17}Department of Physics, Aligarh Muslim University, Aligarh, India
\item \Idef{org18}Department of Physics, Pusan National University, Pusan, Republic of Korea
\item \Idef{org19}Department of Physics, Sejong University, Seoul, Republic of Korea
\item \Idef{org20}Department of Physics, University of California, Berkeley, California, United States
\item \Idef{org21}Department of Physics, University of Oslo, Oslo, Norway
\item \Idef{org22}Department of Physics and Technology, University of Bergen, Bergen, Norway
\item \Idef{org23}Dipartimento di Fisica dell'Universit\`{a} 'La Sapienza' and Sezione INFN, Rome, Italy
\item \Idef{org24}Dipartimento di Fisica dell'Universit\`{a} and Sezione INFN, Cagliari, Italy
\item \Idef{org25}Dipartimento di Fisica dell'Universit\`{a} and Sezione INFN, Trieste, Italy
\item \Idef{org26}Dipartimento di Fisica dell'Universit\`{a} and Sezione INFN, Turin, Italy
\item \Idef{org27}Dipartimento di Fisica e Astronomia dell'Universit\`{a} and Sezione INFN, Bologna, Italy
\item \Idef{org28}Dipartimento di Fisica e Astronomia dell'Universit\`{a} and Sezione INFN, Catania, Italy
\item \Idef{org29}Dipartimento di Fisica e Astronomia dell'Universit\`{a} and Sezione INFN, Padova, Italy
\item \Idef{org30}Dipartimento di Fisica `E.R.~Caianiello' dell'Universit\`{a} and Gruppo Collegato INFN, Salerno, Italy
\item \Idef{org31}Dipartimento DISAT del Politecnico and Sezione INFN, Turin, Italy
\item \Idef{org32}Dipartimento di Scienze e Innovazione Tecnologica dell'Universit\`{a} del Piemonte Orientale and INFN Sezione di Torino, Alessandria, Italy
\item \Idef{org33}Dipartimento Interateneo di Fisica `M.~Merlin' and Sezione INFN, Bari, Italy
\item \Idef{org34}European Organization for Nuclear Research (CERN), Geneva, Switzerland
\item \Idef{org35}Faculty of Electrical Engineering, Mechanical Engineering and Naval Architecture, University of Split, Split, Croatia
\item \Idef{org36}Faculty of Engineering and Science, Western Norway University of Applied Sciences, Bergen, Norway
\item \Idef{org37}Faculty of Nuclear Sciences and Physical Engineering, Czech Technical University in Prague, Prague, Czech Republic
\item \Idef{org38}Faculty of Science, P.J.~\v{S}af\'{a}rik University, Ko\v{s}ice, Slovakia
\item \Idef{org39}Frankfurt Institute for Advanced Studies, Johann Wolfgang Goethe-Universit\"{a}t Frankfurt, Frankfurt, Germany
\item \Idef{org40}Gangneung-Wonju National University, Gangneung, Republic of Korea
\item \Idef{org41}Gauhati University, Department of Physics, Guwahati, India
\item \Idef{org42}Helmholtz-Institut f\"{u}r Strahlen- und Kernphysik, Rheinische Friedrich-Wilhelms-Universit\"{a}t Bonn, Bonn, Germany
\item \Idef{org43}Helsinki Institute of Physics (HIP), Helsinki, Finland
\item \Idef{org44}High Energy Physics Group,  Universidad Aut\'{o}noma de Puebla, Puebla, Mexico
\item \Idef{org45}Hiroshima University, Hiroshima, Japan
\item \Idef{org46}Hochschule Worms, Zentrum  f\"{u}r Technologietransfer und Telekommunikation (ZTT), Worms, Germany
\item \Idef{org47}Horia Hulubei National Institute of Physics and Nuclear Engineering, Bucharest, Romania
\item \Idef{org48}Indian Institute of Technology Bombay (IIT), Mumbai, India
\item \Idef{org49}Indian Institute of Technology Indore, Indore, India
\item \Idef{org50}Indonesian Institute of Sciences, Jakarta, Indonesia
\item \Idef{org51}INFN, Laboratori Nazionali di Frascati, Frascati, Italy
\item \Idef{org52}INFN, Sezione di Bari, Bari, Italy
\item \Idef{org53}INFN, Sezione di Bologna, Bologna, Italy
\item \Idef{org54}INFN, Sezione di Cagliari, Cagliari, Italy
\item \Idef{org55}INFN, Sezione di Catania, Catania, Italy
\item \Idef{org56}INFN, Sezione di Padova, Padova, Italy
\item \Idef{org57}INFN, Sezione di Roma, Rome, Italy
\item \Idef{org58}INFN, Sezione di Torino, Turin, Italy
\item \Idef{org59}INFN, Sezione di Trieste, Trieste, Italy
\item \Idef{org60}Inha University, Incheon, Republic of Korea
\item \Idef{org61}Institut de Physique Nucl\'{e}aire d'Orsay (IPNO), Institut National de Physique Nucl\'{e}aire et de Physique des Particules (IN2P3/CNRS), Universit\'{e} de Paris-Sud, Universit\'{e} Paris-Saclay, Orsay, France
\item \Idef{org62}Institute for Nuclear Research, Academy of Sciences, Moscow, Russia
\item \Idef{org63}Institute for Subatomic Physics, Utrecht University/Nikhef, Utrecht, Netherlands
\item \Idef{org64}Institute for Theoretical and Experimental Physics, Moscow, Russia
\item \Idef{org65}Institute of Experimental Physics, Slovak Academy of Sciences, Ko\v{s}ice, Slovakia
\item \Idef{org66}Institute of Physics, Homi Bhabha National Institute, Bhubaneswar, India
\item \Idef{org67}Institute of Physics of the Czech Academy of Sciences, Prague, Czech Republic
\item \Idef{org68}Institute of Space Science (ISS), Bucharest, Romania
\item \Idef{org69}Institut f\"{u}r Kernphysik, Johann Wolfgang Goethe-Universit\"{a}t Frankfurt, Frankfurt, Germany
\item \Idef{org70}Instituto de Ciencias Nucleares, Universidad Nacional Aut\'{o}noma de M\'{e}xico, Mexico City, Mexico
\item \Idef{org71}Instituto de F\'{i}sica, Universidade Federal do Rio Grande do Sul (UFRGS), Porto Alegre, Brazil
\item \Idef{org72}Instituto de F\'{\i}sica, Universidad Nacional Aut\'{o}noma de M\'{e}xico, Mexico City, Mexico
\item \Idef{org73}iThemba LABS, National Research Foundation, Somerset West, South Africa
\item \Idef{org74}Johann-Wolfgang-Goethe Universit\"{a}t Frankfurt Institut f\"{u}r Informatik, Fachbereich Informatik und Mathematik, Frankfurt, Germany
\item \Idef{org75}Joint Institute for Nuclear Research (JINR), Dubna, Russia
\item \Idef{org76}Korea Institute of Science and Technology Information, Daejeon, Republic of Korea
\item \Idef{org77}KTO Karatay University, Konya, Turkey
\item \Idef{org78}Laboratoire de Physique Subatomique et de Cosmologie, Universit\'{e} Grenoble-Alpes, CNRS-IN2P3, Grenoble, France
\item \Idef{org79}Lawrence Berkeley National Laboratory, Berkeley, California, United States
\item \Idef{org80}Lund University Department of Physics, Division of Particle Physics, Lund, Sweden
\item \Idef{org81}Nagasaki Institute of Applied Science, Nagasaki, Japan
\item \Idef{org82}Nara Women{'}s University (NWU), Nara, Japan
\item \Idef{org83}National and Kapodistrian University of Athens, School of Science, Department of Physics , Athens, Greece
\item \Idef{org84}National Centre for Nuclear Research, Warsaw, Poland
\item \Idef{org85}National Institute of Science Education and Research, Homi Bhabha National Institute, Jatni, India
\item \Idef{org86}National Nuclear Research Center, Baku, Azerbaijan
\item \Idef{org87}National Research Centre Kurchatov Institute, Moscow, Russia
\item \Idef{org88}Niels Bohr Institute, University of Copenhagen, Copenhagen, Denmark
\item \Idef{org89}Nikhef, National institute for subatomic physics, Amsterdam, Netherlands
\item \Idef{org90}NRC Kurchatov Institute IHEP, Protvino, Russia
\item \Idef{org91}NRNU Moscow Engineering Physics Institute, Moscow, Russia
\item \Idef{org92}Nuclear Physics Group, STFC Daresbury Laboratory, Daresbury, United Kingdom
\item \Idef{org93}Nuclear Physics Institute of the Czech Academy of Sciences, \v{R}e\v{z} u Prahy, Czech Republic
\item \Idef{org94}Oak Ridge National Laboratory, Oak Ridge, Tennessee, United States
\item \Idef{org95}Ohio State University, Columbus, Ohio, United States
\item \Idef{org96}Petersburg Nuclear Physics Institute, Gatchina, Russia
\item \Idef{org97}Physics department, Faculty of science, University of Zagreb, Zagreb, Croatia
\item \Idef{org98}Physics Department, Panjab University, Chandigarh, India
\item \Idef{org99}Physics Department, University of Jammu, Jammu, India
\item \Idef{org100}Physics Department, University of Rajasthan, Jaipur, India
\item \Idef{org101}Physikalisches Institut, Eberhard-Karls-Universit\"{a}t T\"{u}bingen, T\"{u}bingen, Germany
\item \Idef{org102}Physikalisches Institut, Ruprecht-Karls-Universit\"{a}t Heidelberg, Heidelberg, Germany
\item \Idef{org103}Physik Department, Technische Universit\"{a}t M\"{u}nchen, Munich, Germany
\item \Idef{org104}Politecnico di Bari, Bari, Italy
\item \Idef{org105}Research Division and ExtreMe Matter Institute EMMI, GSI Helmholtzzentrum f\"ur Schwerionenforschung GmbH, Darmstadt, Germany
\item \Idef{org106}Rudjer Bo\v{s}kovi\'{c} Institute, Zagreb, Croatia
\item \Idef{org107}Russian Federal Nuclear Center (VNIIEF), Sarov, Russia
\item \Idef{org108}Saha Institute of Nuclear Physics, Homi Bhabha National Institute, Kolkata, India
\item \Idef{org109}School of Physics and Astronomy, University of Birmingham, Birmingham, United Kingdom
\item \Idef{org110}Secci\'{o}n F\'{\i}sica, Departamento de Ciencias, Pontificia Universidad Cat\'{o}lica del Per\'{u}, Lima, Peru
\item \Idef{org111}Shanghai Institute of Applied Physics, Shanghai, China
\item \Idef{org112}St. Petersburg State University, St. Petersburg, Russia
\item \Idef{org113}Stefan Meyer Institut f\"{u}r Subatomare Physik (SMI), Vienna, Austria
\item \Idef{org114}SUBATECH, IMT Atlantique, Universit\'{e} de Nantes, CNRS-IN2P3, Nantes, France
\item \Idef{org115}Suranaree University of Technology, Nakhon Ratchasima, Thailand
\item \Idef{org116}Technical University of Ko\v{s}ice, Ko\v{s}ice, Slovakia
\item \Idef{org117}Technische Universit\"{a}t M\"{u}nchen, Excellence Cluster 'Universe', Munich, Germany
\item \Idef{org118}The Henryk Niewodniczanski Institute of Nuclear Physics, Polish Academy of Sciences, Cracow, Poland
\item \Idef{org119}The University of Texas at Austin, Austin, Texas, United States
\item \Idef{org120}Universidad Aut\'{o}noma de Sinaloa, Culiac\'{a}n, Mexico
\item \Idef{org121}Universidade de S\~{a}o Paulo (USP), S\~{a}o Paulo, Brazil
\item \Idef{org122}Universidade Estadual de Campinas (UNICAMP), Campinas, Brazil
\item \Idef{org123}Universidade Federal do ABC, Santo Andre, Brazil
\item \Idef{org124}University College of Southeast Norway, Tonsberg, Norway
\item \Idef{org125}University of Cape Town, Cape Town, South Africa
\item \Idef{org126}University of Houston, Houston, Texas, United States
\item \Idef{org127}University of Jyv\"{a}skyl\"{a}, Jyv\"{a}skyl\"{a}, Finland
\item \Idef{org128}University of Liverpool, Liverpool, United Kingdom
\item \Idef{org129}University of Science and Techonology of China, Hefei, China
\item \Idef{org130}University of Tennessee, Knoxville, Tennessee, United States
\item \Idef{org131}University of the Witwatersrand, Johannesburg, South Africa
\item \Idef{org132}University of Tokyo, Tokyo, Japan
\item \Idef{org133}University of Tsukuba, Tsukuba, Japan
\item \Idef{org134}Universit\'{e} Clermont Auvergne, CNRS/IN2P3, LPC, Clermont-Ferrand, France
\item \Idef{org135}Universit\'{e} de Lyon, Universit\'{e} Lyon 1, CNRS/IN2P3, IPN-Lyon, Villeurbanne, Lyon, France
\item \Idef{org136}Universit\'{e} de Strasbourg, CNRS, IPHC UMR 7178, F-67000 Strasbourg, France, Strasbourg, France
\item \Idef{org137}Universit\'{e} Paris-Saclay Centre d'Etudes de Saclay (CEA), IRFU, D\'{e}partment de Physique Nucl\'{e}aire (DPhN), Saclay, France
\item \Idef{org138}Universit\`{a} degli Studi di Foggia, Foggia, Italy
\item \Idef{org139}Universit\`{a} degli Studi di Pavia, Pavia, Italy
\item \Idef{org140}Universit\`{a} di Brescia, Brescia, Italy
\item \Idef{org141}Variable Energy Cyclotron Centre, Homi Bhabha National Institute, Kolkata, India
\item \Idef{org142}Warsaw University of Technology, Warsaw, Poland
\item \Idef{org143}Wayne State University, Detroit, Michigan, United States
\item \Idef{org144}Westf\"{a}lische Wilhelms-Universit\"{a}t M\"{u}nster, Institut f\"{u}r Kernphysik, M\"{u}nster, Germany
\item \Idef{org145}Wigner Research Centre for Physics, Hungarian Academy of Sciences, Budapest, Hungary
\item \Idef{org146}Yale University, New Haven, Connecticut, United States
\item \Idef{org147}Yonsei University, Seoul, Republic of Korea
\end{Authlist}
\endgroup

\end{document}